\documentclass[conference]{IEEEtran}
\IEEEoverridecommandlockouts
\usepackage{cite}
\usepackage{amsmath,amssymb,amsfonts}
\usepackage{algorithmic}
\usepackage{graphicx}
\usepackage{textcomp}
\usepackage{xcolor}
\def\BibTeX{{\rm B\kern-.05em{\sc i\kern-.025em b}\kern-.08em
    T\kern-.1667em\lower.7ex\hbox{E}\kern-.125emX}}

\usepackage{graphicx}
\usepackage{tabularx}
\newcolumntype{C}{>{\centering\arraybackslash}X}
\usepackage{blindtext}
\usepackage{multirow}  % For multirow table
\usepackage{balance}  % For having balance between the text columns at the end
\usepackage{array} % added for firsthline, multirow, and multicol tables.
\usepackage{booktabs} % added for better tables.
\usepackage{url}  % For dealing with special symbols in url's. In particular in .bib file.
\usepackage{hyperref}  % Added since "URL" package does not allow "_" in the urls.
\usepackage{easyReview}
\usepackage{enumitem}  % To get rid of indentation of itemized list.
\begin{document}

\title{Feature Selection on a Flare Forecasting Testbed:\\A Comparative Study of 24 Methods\\

% ADD YOUR PROPOSED TITLE HERE:
% For AGU Fall Meeting
% Shreejaa:
% Compare 24-Feature subset selection on SWAM_SF
% Krishna:
% Atharv:
% Sagar:

% "20 Feature Selection Algorithms, One Dataset, One Problem: Flare Forecasting"
% "..."

% {\footnotesize \textsuperscript{*}Note: Sub-titles are not captured in Xplore and should not be used}
}

% \author{Anonymous}

\author{
    \IEEEauthorblockN{
        Atharv~Yeolekar\IEEEauthorrefmark{6}\IEEEauthorrefmark{3}\IEEEauthorrefmark{1},
        Sagar~Patel\IEEEauthorrefmark{6}\IEEEauthorrefmark{3}\IEEEauthorrefmark{1},
        Shreejaa~Talla\IEEEauthorrefmark{6}\IEEEauthorrefmark{3}\IEEEauthorrefmark{1},
        Krishna~Rukmini~Puthucode\IEEEauthorrefmark{6}\IEEEauthorrefmark{3}\IEEEauthorrefmark{1}, \\
        Azim~Ahmadzadeh\IEEEauthorrefmark{6}\IEEEauthorrefmark{4}, 
        Viacheslav~M.~Sadykov\IEEEauthorrefmark{2}\IEEEauthorrefmark{4}, and
        Rafal~A.~Angryk\IEEEauthorrefmark{6}\IEEEauthorrefmark{4}
    }
    \IEEEauthorblockA{
        \IEEEauthorrefmark{6}Department of Compute Science,
        Georgia State University, USA \\
        \IEEEauthorrefmark{2}Physics \& Astronomy Department,
        Georgia State University, USA \\
        Email:  \IEEEauthorrefmark{3}\{ayeolekar1,spatel389,stalla1,kputhucode1\}@student.gsu.edu, \\
        \IEEEauthorrefmark{4}\{aahmadzadeh1,vsadykov,rangryk\}@gsu.edu \\
        \IEEEauthorrefmark{1}\textit{These authors contributed equally to this work.}
    }
}

% \and
% \and
% \and
% \and
% \and

% ICDM HDM Workshop: https://www.cs.bham.ac.uk/~axk/HDM21.htm

\maketitle

\begin{abstract}
    % [Azim starts]
    The Space-Weather ANalytics for Solar Flares (SWAN-SF) is a multivariate time series benchmark dataset recently created to serve the heliophysics community as a testbed for solar flare forecasting models. SWAN-SF contains 54 unique features, with 24 quantitative features computed from the photospheric magnetic field maps of active regions, describing their precedent flare activity. In this study, for the first time, we systematically attacked the problem of quantifying the relevance of these features to the ambitious task of flare forecasting. We implemented an end-to-end pipeline for preprocessing, feature selection, and evaluation phases. We incorporated 24 Feature Subset Selection (FSS) algorithms, including multivariate and univariate, supervised and unsupervised, wrappers and filters. We methodologically compared the results of different FSS algorithms, both on the multivariate time series and vectorized formats, and tested their correlation and reliability, to the extent possible, by using the selected features for flare forecasting on unseen data, in univariate and multivariate fashions. We concluded our investigation with a report of the best FSS methods in terms of their top-k features, and the analysis of the findings. We wish the reproducibility of our study and the availability of the data allow the future attempts be comparable with our findings and themselves.
    % [Azim ends]
\end{abstract}


% ----------------------------------
%
%          INTRODUCTION
%
% ----------------------------------
\section{Introduction}\label{sec:introduction}
    With rapid advances in machine learning and its unprecedented success across various domains, in recent years, interdisciplinary researchers have shown growing interest in using such tools and techniques for space-weather analytics and forecast. Extreme space-weather events are somewhat similar to extreme terrestrial events. The difference is that the main creating source of all space-weather events is the Sun. The solar flares triggered in the solar active regions can initiate Coronal Mass Ejections (CMEs) and enhance the flux of Solar Energetic Particles (SEP), both directly affecting the terrestrial environment.
    
    Extreme space-weather events can have severe economic and collateral impact on us and our interconnected technologies here on Earth, as well as in space, due to the growing number of satellites \cite{eastwood2017economic}. This includes Earth climate, aviation and space radiation environment, electric power grid, GPS systems, HF radio communications, satellite communications, and everything that depends on these systems to function properly. In 2008, the direct economic impact of an extreme space-weather event was estimated (by the National Research Council) to be $\$1-2$ trillion for the United States, during the first year alone \cite{nrc:spaceweather}. In realization of such a natural threat, several agencies have been directly studying or funding research in this area, to name a few in the United States alone, the National Science Foundation (NSF), the National Aeronautical and Space Administration (NASA), the National Oceanic and Atmospheric Administration (NOAA), the Department of Defense (DoD), and the Federal Aviation Administration (FAA).
    
    Solar flares are one of such events. They are sudden brightness increases on the Sun visible in the wide wavelength range, from radio to gamma, representing the release of the free magnetic energy in the active region in the form of radiation, heat, kinetic energy of plasma, and accelerated particles. Since 1974, NOAA GOES satellites have been detecting X-ray flares and labeling them based on their peak 1-8\,{\AA} soft X-ray flux. From weakest to strongest flares are classified logarithmically as A ($10^{-8} $ to $10^{-7} W/m^{2}$), B ($10^{-7}$ to $10^{-6} W/m^{2}$), C ($10^{-6}$ to $10^{-5} W/m^{2}$), M($10^{-5}$ to $10^{-4} W/m^{2}$), and X ($>10^{-4} W/m^{2}$), meaning that an X-class flare is 10 times stronger than an M-class flare, and 100 times stronger than a C-class flare. Often in flare forecasting/prediction studies, the magnitude or the probability of occurrence of different classes of flares in an $h$-hour prediction window is of interest, e.g., \cite{barnes2016comparison, bobra2015solar,sadykov2017PIL}.
    
    %%%%%%%%%%%%%%%%%% Slava's two paragraphs
    The vast majority of the implemented approaches in forecasting the occurrence of a strong flare (typically, an M-class flare or larger) utilize properties computed from the photospheric vector or line-of-sight magnetic field maps of solar active regions \cite{bobra2015solar, sadykov2017PIL, nishizuka2017flareforecast}. The SWAN-SF data set \cite{angryk2019multivariate} described in Section~\ref{sec:data_and_preprocessing} contains these characteristics carefully computed for May, 2010~--- December, 2018 time period. Utilization of SWAN-SF for the considered problem of feature ranking in flare forecasting is appropriate and allows to mostly avoid an exhausting data preparation phase.
    
    The main contribution of this paper is the initiation of a series of comparable feature selection strategies for the ambitious task of flare forecasting. Because the existing attempts were carried out on different datasets with different collection strategies, imbalance ratios, and post-processing steps, a fair comparison between the discovered ranks is not possible. Using the flare forecasting benchmark dataset, SWAN-SF, we initiate this attempt that allows future feature selection, ranking, and extraction investigations to be comparable with our findings and with themselves. To help achieve this goal, we do our best to present a reproducible study by extensively explaining the details of data preparation, feature selection, and evaluation strategies and releasing our code \footnote{\url{https://bitbucket.org/gsudmlab/featuresubsetselection_on_swan-sf/}} for further investigation and reusability. This highlights the importance and novelty of the current work. 
    
    This paper is structured as follows: Section~\ref{sec:related} describes the work related to feature ranking, including the attempts for ranking in flare forecasting. Section~\ref{sec:background} summarizes the feature ranking methods utilized in this work. The data and prediction metrics are described in Section~\ref{sec:data_and_preprocessing}. Sections \ref{sec:feature_selection_methodology} and \ref{sec:ranking_evaluation_methodology} highlight the feature ranking and evaluation methodology. The results of the investigation are described in Section~\ref{sec:results} and are followed by conclusions in Section~\ref{sec:conclusion}.

% ----------------------------------
%
%           RELATED WORK
%
% ----------------------------------
\section{Related Work}\label{sec:related}
   %[Shreejaa starts]
   Feature subset selection (FSS) methods are often categorized by their selection strategy from different perspectives. An FSS method is called a \textit{filter} if it searches for an optimal subset based on the general patterns in data regardless of a particular learning model's performance. In contrast, \textit{wrapper} methods utilize a greedy technique to search the space of all possible subset features by using a machine learning algorithm at its core \cite{kohavi1997wrappers}. When an FSS method combines these two strategies, it is then called \textit{embedded} \cite{guyon2003introduction}. The FSS methods which rely on learners can be further divided into two sub-categories: supervised and unsupervised. The former takes advantage of the class labels of the data to find the most relevant features, while the latter group primarily employs clustering algorithms to do so \cite{miao2016survey}. From a different angle, an FSS method may consider multiple features at once to determine their collective relevance, in contrast to single-feature assessment. With this criterion, FSS methods can be classified into \textit{univariate} or \textit{multivariate} groups. Another category of distinction between the FSS methods is the \textit{vectorized} and \textit{MTS-based} algorithms. The former utilizes the descriptive statistics from the multivariate time series data as its input, whereas the latter employs the MTS data in its original format.
   
   Supervised FSS methods employ various supervised machine learning algorithms (also called \textit{estimators} or \textit{base learners} in this context) to identify an optimal subset of the original features by considering the class labels and removing the redundant features (with a minimal or negative impact on the discrimination task) with the a priori knowledge. For example, Support Vector Channel Selection for brain-computer interface (BCI) dataset \cite{lal2004support} combines Recursive Feature Elimination (RFE) \cite{guyon2002gene} and Fisher Criterion (FC) \cite{bishop1995neural} along with Support Vector Machines (SVMs). This supervised wrapper approach established a reliable and intuitive ranking method and quickly became the state-of-the-art FSS method for the proceeding traditional FSS methods. This method however, loses the correlation information as it utilizes a univariate selection. Understanding this shortcoming gave birth to a new family of wrapper methods that let classifiers process some statistical interpretation of data instead of the original data. Corona \cite{yang2005supervised} is an example of such a family. It preserves the correlation information by employing correlation coefficients as features. Due to the general limits of supervised FSS methods \cite{smialowski2010pitfalls}, unsupervised methods are equally useful, or perhaps even more, given that most of the available data are unlabeled. Unsupervised FSS methods, in their simplest form, compute the similarity between the features and then reduce the redundancy in data by dropping the similar features. For example, CLeVer \cite{Yoon2005FeatureSS} selects an optimal subset of features by means of clustering on the loadings obtained by running PCA on the original data.
   
   Univariate FSS methods, as mentioned before, take into account only one feature at a time to identify their relevance to the class label. The Relief algorithm \cite{robnik2003theoretical} is a popular example of this group. Relief is a wrapper, supervised method that determines the relevance of a feature by examining each instance's values with respect to the values of its neighboring instances of the same class. Derived from Relief, a large family of FSS methods has been introduced that particularly attracted the attention in bioinformatics \cite{urbanowicz2018relief}. Equally popular, but in the multivariate category of FSS methods, is the Minimum-Redundancy Maximum-Relevance (mRMR) algorithm that was originally applied to microarray data \cite{peng2005mrmr}. The name is inspired by the core idea behind the algorithm, searching for a subset of features by maximizing their correlation with the class labels and minimizing their correlation with one another. Similar to the Relief algorithm, mRMR also inspired numerous FSS methods \cite{ramirez2017fast}, especially in gene selection studies \cite{akadi2011two}, even mixed with Relief \cite{zhang2007two}.
   
   Among the wrapper methods, the Support Vector Channel Selection algorithm (that we mentioned earlier) is perhaps the most popular one. This FSS method ranks features by training an SVM on the labeled data and assigning scores to features based on the concept of margin maximization \cite{guyon2002gene}, and was originally applied for the selection of electroencephalogram (EEG) channels. Among the filter FSS methods, the F-score \cite{ding2009feature}, mutual information \cite{vinh2012novel}, and information gain \cite{azhagusundari2013feature} methods laid out the basic selection ideas that feed many complex filter methods. The Fast Correlation-Based Filter (FCBF) method \cite{yu2003feature} is another popular method in this category.
   
   % Slava's paragraph
   Although not widely used, the application of feature ranking methods for flare forecasting is definitely not a new idea. One of the most frequently used feature ranking techniques is the univariate F-score \cite{bobra2015solar, sadykov2017PIL, cinto2020extremegrad}. Another popular FSS methodology mentioned in the research papers employing a Random Forest algorithm \cite{breiman2001randomforest} for flare forecasting are a selection of features based on Gini importance or the Mean Decrease Gini \cite{liu2017rfflare}. Other examples are the use of unsupervised feature extraction/selection from the magnetic Polarity Inversion Line (PIL) masks based on kernel PCA \cite{wang2020kernelPCA}, and the employment of CBF- and mRMR-like FSS methods \cite{ahmed2013asap}. Nevertheless, to the best of our knowledge, the effectiveness of a variety of other FSS methods on flare forecasting datasets was never examined. Our objective is to study the efficacy of a larger number of methods, but more importantly, to introduce a framework of comparability for future attempts.
   
   It is worth noting that in the literature, some times the terms \textit{selection}, \textit{ranking}, and \textit{extraction} are used loosely and even interchangeably. To avoid confusion, we clarify our understanding of these terms here before we move on to the next section. \textit{Feature extraction} is specific to the algorithms which create (extract) new features by combining the original ones. The objective in this family of methods is to find a new feature space such that it has a lower dimension than the original space, and at the same time, allows a better separability of classes. PCA \cite{pearson1901liii} is the typical example of a feature extraction method. Each new feature extracted by PCA is a linear combination of all the original features (possibly with close-to-zero coefficients for some of them). While this approach provides more flexibility as it is not limited to the original feature space, for problems where the importance of the existing features is of interest, it is not useful. This is where \textit{feature selection} methods come into play. These methods aim at finding an optimal subset of the original features in such a way that the selected features carry most of the information about the distribution of data while excluding the redundant features, i.e., highly correlated or noisy features. The term \textit{feature subset selection} (FSS) is sometimes used to emphasize this difference between selection and extraction. A sub-family of this group of methods additionally assign a score to each feature and find the optimal subset by means of thresholding. They are known as the \textit{feature ranking} methods.
%   \cite{pearson1901liii, wold1987principal} ---> removed due to space limit.

   %[Shreejaa ends]
% ----------------------------------
%
%           BACKGROUND
%
% ----------------------------------
\section{Background}\label{sec:background}
    
    A summary of the methods we employed in this study, in order to gain insight into the relevance of the SWAN-SF's features, is given in Table.~\ref{tab:all_fss}. In our list, there are filters and wrappers, supervised and unsupervised methods, univariate and multivariate approaches, and vectorized-based and Multivariate Time Series-based (MTS-based) strategies. In this section, we briefly review their implementations.
     
    %[Rukmini starts]
    \begin{table*}[h!]
    \setlength{\extrarowheight}{2pt}
    \centering
    % \begin{tabularx}{0.863\textwidth}{|c | c | c | c | c | c | c | c | c | c | c | c | c|} 
    \begin{tabularx}{0.95\textwidth}{|c | c | c | c | c | c | c | c | c | c | c | c | c|} 
    \hline
    &\textbf{Abbreviation} & \textbf{FSS Algorithm} & \textbf{Estimator/Method} & \rotatebox[origin=c]{-90}{\textbf{Supervised}}  &\rotatebox[origin=c]{-90}{\textbf{Unsupervised}} & \rotatebox[origin=c]{-90}{\textbf{Multivariate}}  &\rotatebox[origin=c]{-90}{\textbf{Univariate}}  & \rotatebox[origin=c]{-90}{\textbf{Vectorized}}  &\rotatebox[origin=c]{-90}{\textbf{MTS}} & \rotatebox[origin=c]{-90}{\textbf{Embedded}}  &\rotatebox[origin=c]{-90}{\textbf{Wrapper}} & \rotatebox[origin=c]{-90}{\textbf{Filter}} \\\hline\hline
    1 & SFM\_Logistic &  \multirow{7}{*}{Select From Model} & Logistic Regression & \checkmark &  & \checkmark &  & \checkmark &  & \checkmark &  & \\ %clear
    2 & SFM\_SVC &  & SVM & \checkmark &  & \checkmark &  & \checkmark &  & \checkmark &  &  \\  %clear
    3 & SFM\_RF &  & Random Forest & \checkmark &  & \checkmark &  & \checkmark &  & \checkmark &  &  \\  %clear
    4 & SFM\_ADA &  & Ada Boost & \checkmark &  & \checkmark &  & \checkmark &  & \checkmark &  &  \\  %clear
    5 & SFM\_Gb &  & Gradient Boosting& \checkmark &  & \checkmark &  & \checkmark &  & \checkmark &  &  \\  %clear
    6 & SFM\_BAG &  & Bagging Trees& \checkmark &  & \checkmark &  & \checkmark &  & \checkmark &  &  \\  %clear
    7 & SFM\_XT & & Extremely Randomized Trees & \checkmark &  & \checkmark &  & \checkmark &  & \checkmark &  &  \\ \hline %clear
    8 & RFE\_RF & \multirow{7}{*}{Recursive Feature Elimination} & Random Forest & \checkmark &  & \checkmark &  & \checkmark &  &  & \checkmark &  \\ %clear
    9 & RFE\_ADA &  & Ada Boost & \checkmark &  & \checkmark &  & \checkmark &  &  & \checkmark &   \\  %clear
    10& RGE\_Gb &  & Gradient Boosting & \checkmark &  & \checkmark &  & \checkmark &  &  & \checkmark &   \\  %clear
    11& RFE\_BAG &  & Bagging Trees & \checkmark &  & \checkmark &  & \checkmark &  &  & \checkmark &   \\   %clear
    12& RFE\_XT &  & Extremely Randomized Trees & \checkmark &  & \checkmark &  & \checkmark &  &  & \checkmark &   \\   %clear
    13& RFE\_SVC &  & SVM & \checkmark &  & \checkmark &  & \checkmark &  &  & \checkmark &   \\  %clear
    14& RFE\_logistic &  & Logistic Regression & \checkmark &  & \checkmark &  & \checkmark &  &  & \checkmark &   \\  \hline %clear
    15& SKB\_MI & \multirow{3}{*}{Select $k$-Best} & Mutual Information & \checkmark & &  & \checkmark & \checkmark &  &  &  & \checkmark\\   %clear
    16& SKB\_FVAL &  & F-statistic & \checkmark &  &  & \checkmark & \checkmark &  &  &  & \checkmark  \\   %clear
    17& SKB\_CHI &  & Chi Square & \checkmark &  &  & \checkmark & \checkmark &  &  &  & \checkmark \\  \hline %clear
    18 & mRMR & Min Redundancy Max Relevance & F-statistic / Pearson Correlation & \checkmark &  & \checkmark &  & \checkmark &  &  &  & \checkmark\\  \hline %clear
    19& Relief & Relief & F-statistic / Pearson Correlation & \checkmark &  & & \checkmark & \checkmark &  &  &  & \checkmark \\ \hline %clear
    20& Corona & Correlation as Features & SVM & \checkmark & & \checkmark &  & \checkmark &  &  & \checkmark &\\  \hline %clear
    21& Clever & CLeVer & $k$-means / PCA &  & \checkmark & \checkmark &  &  & \checkmark & \checkmark &  & \\  \hline %clear
    22& PIE & PIE &  Normalized Mutual Information & \checkmark &  & \checkmark &  &   & \checkmark &   &  & \checkmark\\  \hline %clear
    23& CSFS & CSFS & Pairwise Mutual Information & \checkmark &  & \checkmark &  &   & \checkmark &   &  & \checkmark\\ \hline %clear
    24& FCBF & FCBF & Mutual Information & \checkmark &  & & \checkmark &   & \checkmark &   &  & \checkmark\\ %clear
    \hline %clear
    \end{tabularx}
    \vspace{0.1cm}
    
    \caption{\label{tab:all_fss} List of all FSS methods used in this study, with their specifications for reference}
    \vspace{-0.7cm}
    \end{table*}

    %[Rukmini Ends]
    
    %[Shreejaa starts]
    \noindent \textbf{Recursive Feature Elimination (RFE)} \cite{guyon2002gene} is a feature selection method that fits data using a base learner such as Random Forest or Logistic Regression, and removes the weakest feature(s) recursively until the stipulated number of features is reached. Either the model's coefficients or the feature importance attributes are used to rank the features. Recursively eliminating features, RFE attempts to eliminate dependencies and collinearity in the model (if any).
     
    % \noindent \textbf{Recursive Feature Elimination (RFE):} Recursive Feature Elimination is a feature subset selection method proposed by Guyon et al. \cite{guyon2002gene} based on SVM. RFE trains the MTS data on SVM to assign \add{the} weights \replace{for}{to} each feature\replace{; based on those weights, ranks to the features are set}{The features are then ranked directly based on the assigned weights}. The lowest-ranked features are dropped \replace{for}{at} each iteration. This process is iterated until the required number of features are retained. All the selected features are given rank 1. Here we set (no of features to select) as 1 which continues the recursive elimination till only 1 feature is left thereby giving unique ranking to all features\comment{.}{@Shreejaa: I could not follow the logic here. If weights are used for ranking, then why selected features are assigned 1? And then, how this will be uniquely ranked at the end?! Please modify this.}
   
   \noindent \textbf{Correlation as Features (Corona)} \cite{yang2005supervised} operates by computing the correlation coefficients matrix $C$ for each MTS instance. The feature selection takes place on the upper triangular of $C$. Using RFE with SVM as the estimator, each feature is assigned a weight. The weights are then used to rank the features in ascending order to generate a rank vector $RV$. A symmetric matrix $SM$ is generated by shaping $RV$ as the lower and upper triangular matrix with diagonals as zero. To obtain ranks of original features, the column-wise summation is applied to $SM$.
   
   \noindent \textbf{Relief} \cite{robnik2003theoretical} gauges the quality of each feature by estimating the separability it provides for distinguishing the neighboring instances of different classes. The algorithm randomly selects an instance $R$ and finds its nearest neighbors $H$ from the same class (i.e., hit) and $M$ from the opposite class (i.e., miss). The feature score for an arbitrary feature $f$ is initialized with zero and updated by $W[f]:= W[f] - \frac{d_{f}(R_i,H)}{m} + \frac{d_{f}(R_i,M)}{m}$, where $d_{f}$ is the Manhattan distance between the feature $f$ of two instances, and $m$ is the longest distance in the space and is used for the normalization purpose. It can be observed that the feature score is (a) penalized for long distances between instances of similar class, and (b) incentivized for long distances between opposite class. This approach claims to have contextual awareness and performs well in problems with strong dependencies between the features.
   
   \noindent\textbf{Minimum Redundancy Maximum Relevance (mRMR)} \cite{peng2005mrmr} finds an optimal subset of feature by iteratively adding features with maximum relevance with the class labels while having minimum redundancy with other already-selected features. The relevance is determined using ANOVA test with respect to the class labels, and the redundancy is determined by mean Pearson correlation in accordance with the selected features. For every $i$-th iteration, mRMR selects the feature having the highest score determined by $\text{score}_i(f) = \frac{F(f,\text{class})}{\sum_{s\in \mathcal{S}}|r(f,s)|/(i-1)}$. In this formula, $f$ is an arbitrary feature, $F$ is the $F$-statistic, $r$ denotes the Pearson correlation, and $\mathcal{S}$ is the set of all features selected prior to $f_i$. This approach is known to be computationally efficient and robust to different downstream classification models \cite{zhao2019maximum}.

   \noindent \textbf{Select $k$-Best} \cite{scikit-learn} is a simple univariate feature selection approach that utilizes a scoring function passed as a parameter. Various statistical tests such as ANOVA test and Chi-Square test, and mutual information can be used to quantify the scoring function. The scoring function must return an array of scores corresponding to all features. Select $k$-Best then retains the top $k$ features with the highest scores. 
   
   \noindent \textbf{Select From Model} \cite{scikit-learn} determines the feature importance based on the estimators' optimization weights, e.g., the \textit{coefficient}s in Logistic Regression or \textit{mean reduction} in the Gini index. The ones with scores lesser than a pre-set threshold parameter are excluded and considered unimportant. The top $k$ features are those which obtained the highest scores.
   %[Shreejaa ends]
   
   % @Krishna
   
   %[Krishna Starts]
   \noindent \textbf{CLeVer} \cite{Yoon2005FeatureSS} is a family of unsupervised FSS methods for MTS data based on descriptive common principal component analysis (DCPC). DCPCs are obtained by bisecting the angles between their principal components after each MTS instance undergoes PCA. The correlation matrix of each MTS instance is passed as an input to obtain its correlation information. The CLeVer algorithm comprises three phases: (1) Computing the Principal Components for each MTS instance by applying Singular Value Decomposition on the correlation matrix; (2) Computing the DCPDs which depict the common direction of the maximum variance of all MTS instances; (3) Ranking of the features by their contribution to the DCPC model features. The last step is carried out by applying the $\ell^2$-norm on the DCPCs (CLeVer-Rank), which is followed by running the $k$-means clustering algorithm to eliminate the redundant features (CLeVer-Cluster).

   \noindent \textbf{Fast Correlation-Based Filter (FCBF)} \cite{yu2003feature}, instead of a traditional linear correlation, employs Symmetric Uncertainty (SU), which involves entropy-based information gain (aka mutual information). To recap, the entropy of a random variable in information theory is the average level of uncertainty information inherent in the variable's possible outcomes. As a result, the amount by which a feature's entropy decreases reflects additional information about features provided by the label and is referred to as information gain. Since this information gain is biased in favor of features with more values, the data must be normalized beforehand. In FCBF, this normalization is accomplished by the Symmetrical Uncertainty, by first determining the probability distribution of each feature, and then its entropy corresponding to its class label. FCBF compiles a list of relevant features based on a given threshold and sorts them by their Symmetric Uncertainty values in descending order. 
   
   % atharv paper 1
   \noindent \textbf{Power Iteration Embedding (PIE)} \cite{han2005supervised} ranks the features to compute the similarity graph for all features (time series) across all time segments. Dynamic Time Warping (DTW) is used to calculate the distance between time series \cite{Serra2014journal}. The largest eigenvector of the normalized adjacency matrix of the graph is computed to determine the cluster structure of these segments. Features' relevance score is evaluated using Normalized Mutual Information \cite{shi2000normalized} between the eigenvector and the ground truth labels. Since the ranks of features with high correlation tend to be similar, to avoid redundancy, a subset selection algorithm is employed. The first step in the subset selection algorithm is to calculate the adjacency matrix of each sensor and then find a linear combination of these matrices that approximates the similarity matrix of the labels while using a small number of minimally redundant sensors.

   % atharv paper 2
   \noindent \textbf{Class Separability Feature Selection (CSFS)} \cite{HAN201329} relies on Mutual Information (MI) which measures both linear and non-linear correlations between the features. Since the pairwise MI matrix is symmetric, an MTS instance is vectorized to build new features using the upper triangular matrix. To rank these features, the ratio of the between-class scatter matrix to the within-class scatter matrix is considered, which is computed over the MI matrix. To avoid redundancy, the between-feature scatter matrix, which reflects the relation between features in terms of class labels, is produced. The larger the matrix value, the lesser the redundancy between its features.

% ----------------------------------
%
%           DATA AND METRICS
%
% ----------------------------------
% [Atharv Starts]
\section{Data and Metrics}\label{sec:data_and_preprocessing}

    \subsection{Data}        
        The Space-Weather ANalytics for Solar Flares (SWAN-SF) benchmark dataset \cite{angryk2019multivariate} comprises multivariate time series data in five classes, namely X, M, C, B, and a non-flaring class denoted by FQ (flare quiet). The dataset is broken down into five temporally non-overlapping partitions such that each partition has the same number of X- and M-class flares. The various time series parameters of the dataset are derived from the solar photospheric magnetograms and NOAA's flare history. Magnetograms and their metadata are provided by the Solar Dynamics Observatory's HMI Active Region Patches (HARP) data product. To explore the details of collection, integration, and curation of the SWAN-SF benchmark dataset, as well as the definitions of the physical parameters we aim to rank in this study, we encourage the reader to see the corresponding publication. The dataset is also publicly available on the Harvard Dataverse repository \cite{DVN/EBCFKM_2020}.
        
    \subsection{Metrics}
        In this paper, the metrics utilized as performance measures are the True Skill Statistic (TSS) \cite{hanssen1965relationship} and the updated Heidke Skill Score (HSS) \cite{balch2008updated} (sometimes denoted as HSS2). These are the two metrics that domain experts found appropriate for the task of flare forecasting to provide a meaningful evaluation of models given the scarcity of strong flares \cite{barnes2008evaluating, Bloomfield_2012}. TSS measures the difference between the true positive rate (detection) and the false positive rate (false alarm). In other words, TSS = $\frac{t p}{p}-\frac{f p}{n}$; where $p=t p+f n$ and $n=f p+t n$. It ranges from -1 to +1, with -1 indicating that the model makes all the wrong predictions/classifications, 0 implying that the model possesses no skill (random-guess), and +1 representing a model that assigns correct class labels to all the instances. TSS is not sensitive to class imbalance \cite{ahmadzadeh2021how, bobra2015solar}. This allows comparison of models which are trained on subsets of data with different imbalance ratios. HSS measures the performance of a model by comparing it to a random-guess model. It is formulated as $\frac{2((t p \cdot t n)-(f n \cdot f p))}{p(f n+t n)+n(t p+f p)}$. HSS2 ranges from -1 to +1, where 0 intimates no difference between the model's performance and random-guess. The magnitude of the negative value is directly proportional to the number of misclassified samples. The positive values quantify how much the model performs better than the random-guess model.

% % ----------------------------------
% % ----------------------------------
% %
% %
% %

    \begin{figure*}
        \centering
        \centering\includegraphics[width=\linewidth]{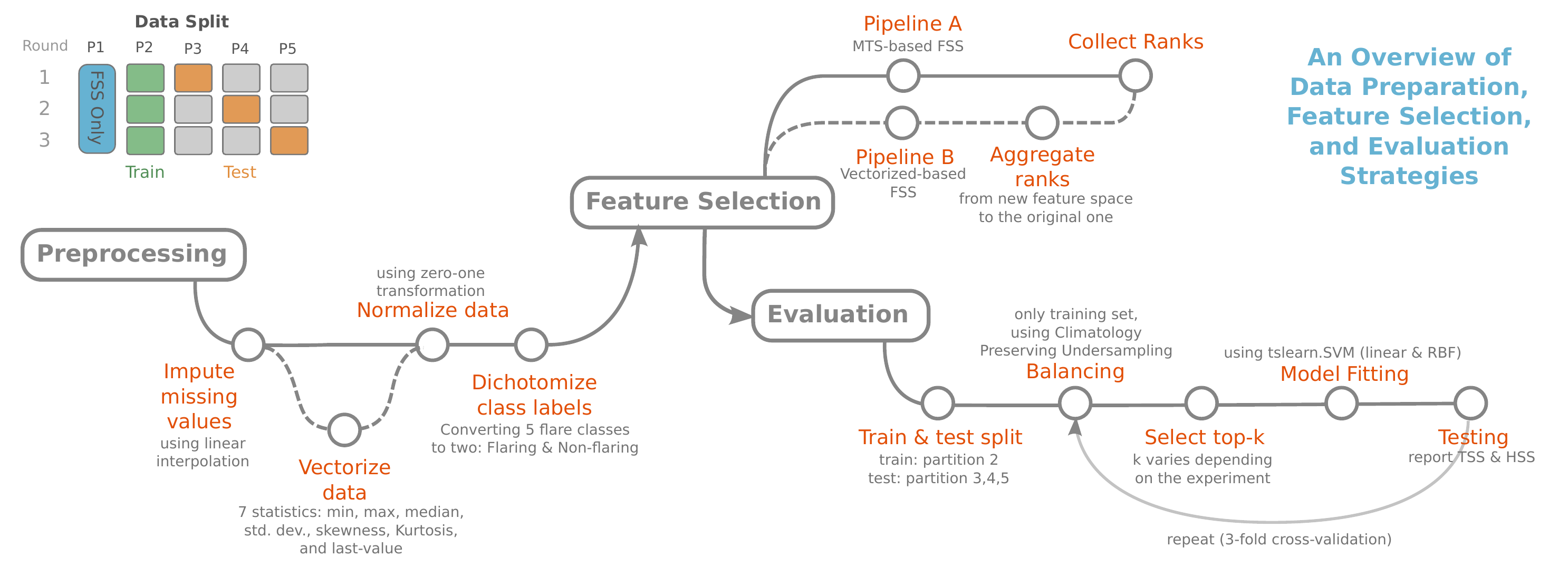}
        \caption{The schematic diagram illustrating the 3 main components of our study: data preprocessing, feature selection, and evaluation. Each part is explained in details in their corresponding sections. The dashed lines indicate the process that is only applicable to vectorized data. The usage of SWAN-SF's 5 partitions is also shown on the top-left area.}
        \label{fig:methodology}
        \vspace{-0.5cm}
    \end{figure*}

\section{Feature Selection Methodology}\label{sec:feature_selection_methodology}
    The list of all FSS methods we utilize in this study, along with their specifications, is given in Table~\ref{tab:all_fss}. We set up two feature selection pipelines, one for vectorized-based FSS methods and the other for MTS-based methods. In this section, we provide a general overview of the pre-processing steps, as well as the strategies adopted in our feature ranking methodology. Fig.~\ref{fig:methodology} is provided to aid in the following of these steps.
    
    % We investigate two general approaches to explore the relevance of features present in the SWAN-SF benchmark dataset; one is to preprocess the MTS data directly and feed them into SVM, and the other is to vectorize the data by extracting a set of descriptive statistics from the time series, followed by training of SVM on these descriptors.
        
    \subsection{Data Preparation}
        The first 24 physical parameters of SWAN-SF are quantitative features describing the magnetic fields in active regions, and we limit our study only to them. To prepare our data for both the ranking and evaluation, first, we impute the missing values, which account for $\leq0.01\%$ of the entire data. We linearly interpolate the missing values using their neighboring values. For the boundary values, we use backward and forward fill, i.e., propagating the last valid values forward or backward.
        
        As majority of the FSS methods we utilize are designed for tabular data (as opposed to MTS data), we need to create a vectorized version of SWAN-SF as well. To do so, we compute 7 descriptive statistics for each physical parameter, yielding 168 statistical features. These statistics are \textit{min}, \textit{max}, \textit{median}, \textit{standard deviation}, \textit{skewness}, \textit{kurtosis}, and \textit{last value} which is simply the last value of each time series. These basic statistics have been used before in several studies working on SWAN-SF to simplify the problem \cite{ahmadzadeh2021how, hostetter2019understanding, ahmadzadeh2019rare}. 
        
        As mentioned in Section~\ref{sec:data_and_preprocessing}, SWAN-SF is a collection of 5 different flare types. For simplification of the problem, we group the stronger flares (X- and M-class flares) and weaker flares (C- and B-class flares, plus FQ class) and create a dichotomized dataset.

    \subsection{Rank Aggregation for Vectorized Data}
        Since the features in the vectorized data are the descriptive statistics and not the original SWAN-SF features, in order to be able to transfer the obtained ranks back to the original feature space, we assign each feature the average scores obtained across all 7 statistics corresponding to that features. This way, we obtain a single score per (original) feature. We use this score, similar to the scores MTS-based FSS methods return, to rank the features.
        
    For the preparation of data, extraction of statistical features, and model fitting we used the pandas and scikit-learn \cite{scikit-learn}, MVTS Data Toolkit \cite{mvts2020ahmadzadeh}, and tslearn \cite{tavenard2020tslearn} packages, respectively.

\section{Evaluation Methodology}\label{sec:ranking_evaluation_methodology}
    A common way to validate the selected features is through domain knowledge and expertise. However, such knowledge does not always exist, and more often than not, it happens that experts have contrasting opinions. In this section, we investigate the reliability of the obtained rankings from three different angles; using an independently trained and tested classifier in a univariate and multivariate fashion, and by looking at the correlations of the rankings across all the 24 FSS methods we employed.

    \subsection{Pre-Evaluation: Sampling and Balancing}\label{subsec:sampling_balancing}
        A dataset is class imbalanced when the frequency of samples of one or more data classes are significantly smaller than those of the majority classes. This is known as the class-imbalance issue \cite{kubat1997addressing} and results in superficial performance if not treated properly. The SWAN-SF benchmark dataset \cite{angryk2019multivariate} exhibits a severe case of class imbalance across all the partitions. To remedy this issue, we employ the \textit{climatology-preserving undersampling} method as suggested in \cite{ahmadzadeh2021how}. This is a stratified undersampling (of all 5 flare classes) where the obtained samples constitute a 1:1 ratio between the strong and weak flares, i.e., the total number of X and M instances is the same as the total number of C, B, and FQ instances. Using this strategy, the subclass ratios will be preserved, hence the name. Note that this strategy is only used for the evaluation phase and only during the model training, and not for testing, i.e., the test set is not balanced. Also, the FSS methods had access to the entire Partition 1 and not subsets of it.
        
        Training and testing on the same partition of SWAN-SF results in unreliable performance because of the auto-correlation of the temporally neighboring data points within the partition, and hence it breaks the underlying assumption that random variables need to be independent for models. \cite{ahmadzadeh2021how} studied this notion on SWAN-SF and defined it as \textit{temporal coherence} of data. Following their suggestion, as shown in Fig.~\ref{fig:methodology}, we avoid mixing partitions for different purposes. Partition 1 is used only to compute the FSS ranks; Partition 2 is reserved for training SVM on the ranked features; a 3-fold cross-validation, each fold being repeated 5 times to generate different undersampled versions of the training partition. The uncertainties are thus computed using the 15 scores. Partitions 3, 4, and 5 are assigned to testing the efficacy of the obtained ranked using SVM.

    \subsection{Univariate Ranking Evaluation}\label{subsec:univariate_evaluation}
        % Univariate feature ranking examines each feature individually to ascertain the strength of the feature's relationship with the response variable.
        Our univariate evaluation pipeline is designed to assess the quality of ranking of features individually, regardless of the ranking methodology (univariate or multivariate). This pipeline functions as follows: it create a subset of SWAN-SF, that contains only one feature and the class labels. It uses Partition 2 of this subset as the training set, and Partitions 3, 4, and 5 as the test set. Note that for obtaining the ranked features, only Partition 1 was used. Next, as shown in Fig.~\ref{fig:methodology} (the evaluation branch), it carries out a 3-fold cross-validation (repeating each trial 5 times) during which we train SVM on subsets of Partition 2 and test it on the test set. This subset is obtained at each iteration, using the climatology-preserving undersampling method discussed in Section~\ref{subsec:sampling_balancing}. The model performance is reported in terms of the average TSS and HSS, with the variability being reported by their standard deviation. The pipeline repeats this procedure for the remaining features and reports the results similarly.
    
    \subsection{Multivariate Ranking Evaluation}\label{subsec:multivariate_evaluation}
        Our multivariate evaluation methodology is implemented to examine the collective contribution of the selected features, regardless of the ranking methodology (univariate or multivariate). The pipeline is very similar to that of the univariate, with one difference that instead of a single feature, a collection of the features is examined at each iteration, starting with one single feature (the best one according to a given FSS method), and iteratively appending the next best feature (according to the same FSS method) to the collection. Since there are $2^{24}$ possible subsets, unlike the univariate pipeline, this has to be repeated for each FSS method's returned ranks separately.

    \subsection{Cross Comparison of Ranked Features}\label{subsec:cross_comparison_evaluation}
        To study the degree of which all the employed FSS methods agree with each other's rankings, we compare all returned rankings and measure their Pearson correlation coefficient. Higher positive correlations indicate that the features are similarly ranked by the compared methods, and conversely, the higher negative correlations indicate the opposite order of relevance. Values closer to zero should be interpreted as no (linear) correlation. While some degree of positive correlation is expected, it is important to note that such a comparative analysis on its own is not indicative of the efficacy of the obtained rankings.

    Note that, although we have two groups of FSS methods, namely vectorized-based and MTS-based, in our evaluation pipeline, we only use MTS data (without vectorization). This decision is made for two reasons: (1) to guarantee the comparability of the algorithms and their rankings, and (2) to test rankings' efficacy on the raw time series, since the statistics chosen for vectorization of the MTS data can change from one study to another. Also, for the sake of completeness, we run all of our training experiments once with the \textit{rbf} kernel and then with the \textit{linear} kernel. We do not use \textit{gak} kernel however, because (1) we did not observe any significant improvement (compared to the other two) in our studies on subsets of the data, despite the theoretical support that the \textit{gak} might be more appropriate for the high-dimensional data such as ours. Also, (2) it is extremely resource hungry due to its reliance on the optimization needed for DTW. We suspect that the ineffectiveness of the \textit{gak} kernel on SWAN-SF might be rooted in the noisiness of the data. Smoothing the SWAN-SF's time series requires a rigorous and comprehensive analysis that we consider out of the scope of this work.
        
    % [Atharv ends]
% ----------------------------------
%
%       RESULTS
%
% ----------------------------------
\section{Results}\label{sec:results}
    In this section, we analyze the output of the pipelines illustrated in Fig.~\ref{fig:methodology}, through multiple lenses. For description of the SWAN-SF feature names and their physical meaning, please see Table 1 in \cite{angryk2019multivariate}.
    
    \begin{figure}
        \centering
        \centering\includegraphics[width=\linewidth]{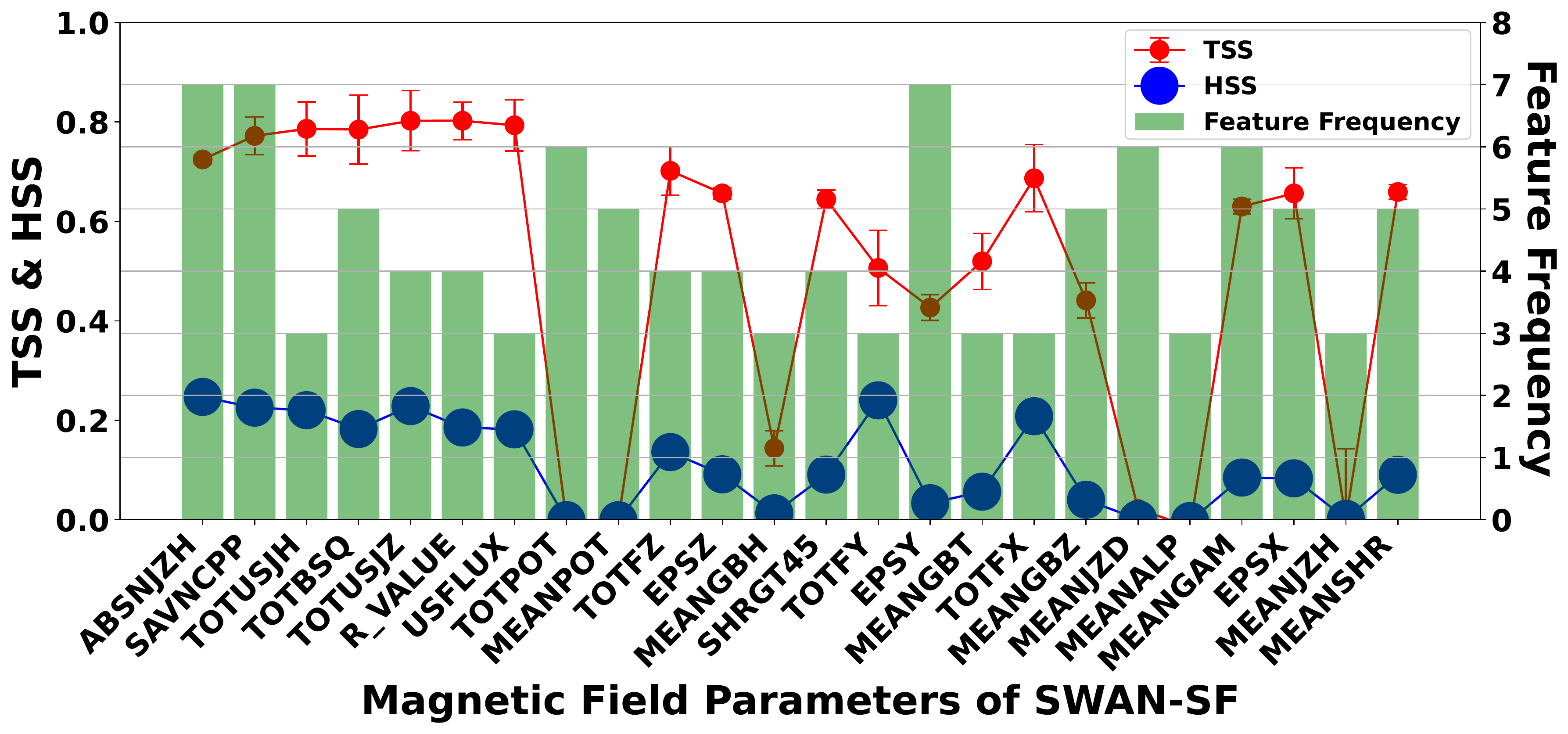}
        \caption{Univariate comparison of the ranked features: On the $x$-axis the 24 SWAN-SF features are listed and sorted by their aggregated ranks across all FSS methods. The green bars represent features' frequency of occurrence, and their position shows their overall ranks across all 24 FSS methods. TSS and HSS are reported as models' performance.}
        \label{fig:univariate_ranks_tss_hss}
        \vspace{-0.4cm}
    \end{figure}
    
    % Univariate
    The result of our univariate evaluation is illustrated in Fig.~\ref{fig:univariate_ranks_tss_hss}. The bar at the $i$-th position represents how often the corresponding feature was ranked as the $i$-th feature. For example, \textit{TOTBSQ} is the 4th most relevant feature as it was overall ranked 5 times (highest) as the 4th feature. While there seems to be a decreasing trend in terms of TSS and HSS, the high fluctuations prevent us from drawing any conclusions in terms of the reliability of the ranking. The first seven features (\textit{ABSNJZH}, \textit{SAVNCPP}, \textit{TOTUSJH}, \textit{TOTBSQ}, \textit{TOTUSJZ}, \textit{R-VALUE}, and \textit{USFLUX}) correspond to the highest individual TSS values, with HSS values being among the highest. On the other hand, even features from the bottom of the list, e.g., \textit{MEANSHR} and \textit{EPSX}, achieve very high TSS values, although accompanied by low HSS values. We know that a good flare forecast performance corresponds to comparable TSS and HSS values \cite{ahmadzadeh2021how}. Also, it is interesting to note that features like \textit{TOTPOT} and \textit{MEANPOT} which correspond to near-zero TSS and HSS, are ranked individually higher than other features such as \textit{TOTFX} which corresponds to a much higher TSS ($\sim0.7$) value. This suggests that while some features (e.g., \textit{TOTPOT}) may not be found relevant individually, when combined with others they might help better discriminate the strong and weak flares. This is why multivariate selection strategies are often preferred over the univariate ones, given their limitations. Interestingly, the top-10 SWAN-SF features according to their aggregated ranks across all FSS methods correspond to the top-11 features for flare prediction mentioned by \cite{bobra2015solar} according to the F-score (the authors also had the \textit{AREA\_ACR} within the top-11 features which is not a part of SWAN-SF). 
    
    \begin{figure}
        \centering
        \centering\includegraphics[width=\linewidth]{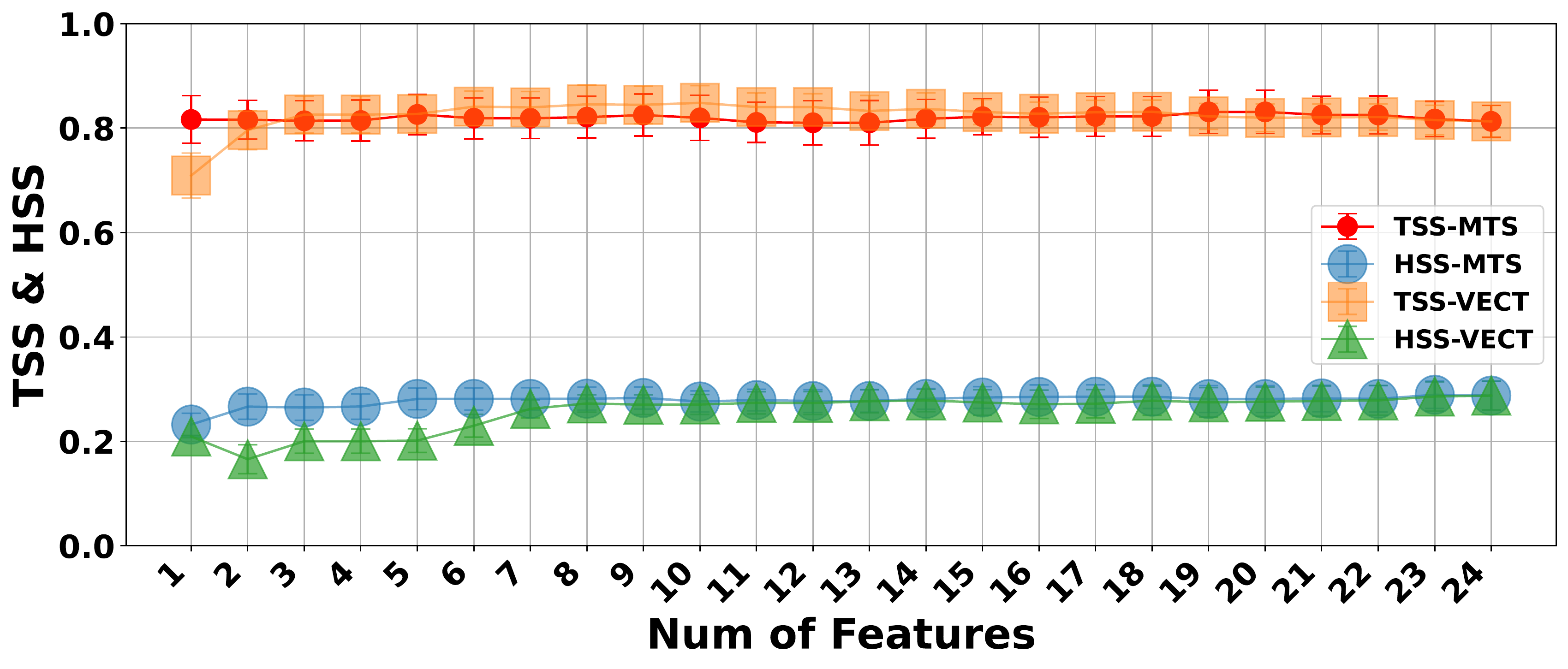}
        \caption{Multivariate comparison of the ranked features: The $x$-axis represents $k$ in the top-$k$ features of SWAN-SF ranked by the mean rank of different MTS- and vectorized-based methods. The $y$-axis represents the mean TSS and mean HSS across all 24 methods. The \textit{linear} kernel is used for this set of experiments.}
        \label{fig:linear_mvts_vect_tss_hss}
        \vspace{-0.4cm}
    \end{figure}
    
    \begin{figure}
        \centering
        \centering\includegraphics[width=\linewidth]{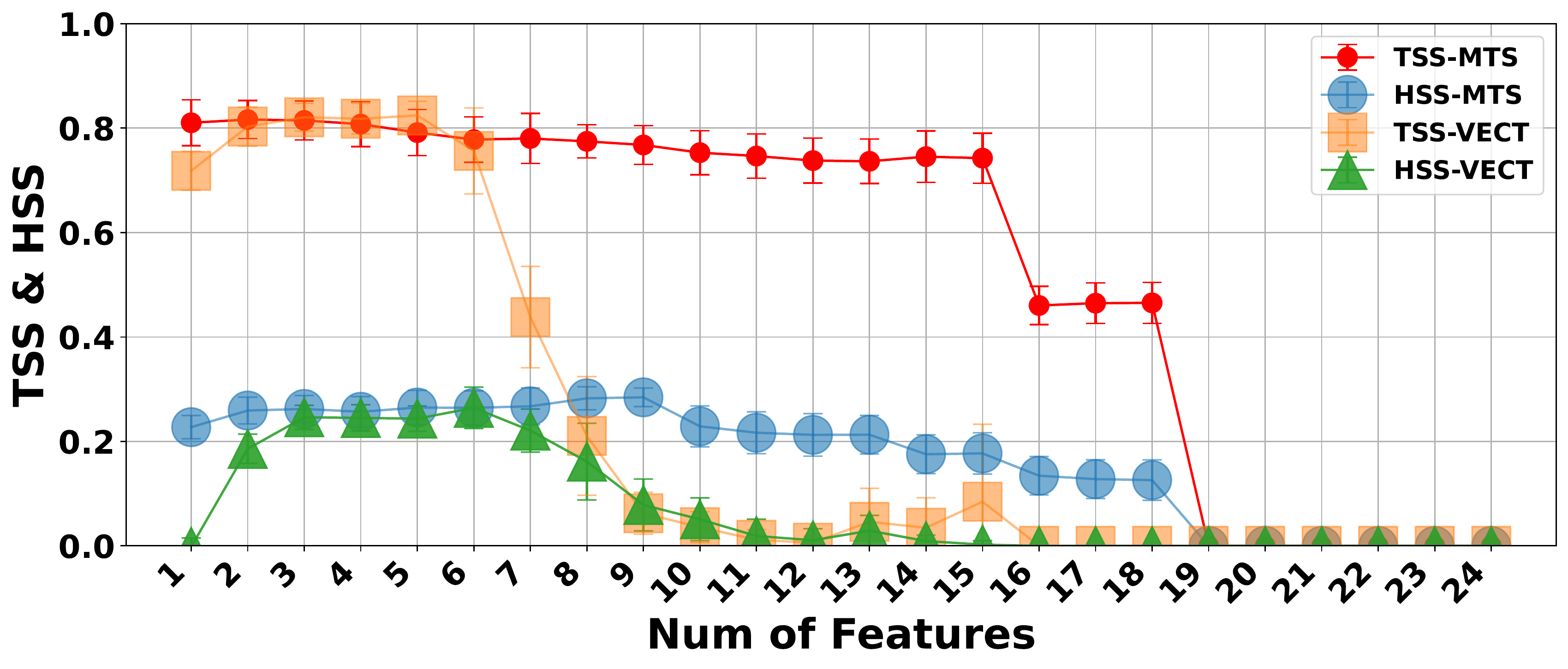}
        \caption{Multivariate comparison of the ranked features: This is similar to Fig.~\ref{fig:linear_mvts_vect_tss_hss}, except that the \textit{rbf} kernel is used (instead of \textit{linear}) for this set of experiments.}
        \label{fig:rbf_mvts_vect_tss_hss}
        \vspace{-0.4cm}
    \end{figure}
    
    % Multivariate
    The results of our multivariate evaluation are illustrated in Figs.~\ref{fig:linear_mvts_vect_tss_hss} and \ref{fig:rbf_mvts_vect_tss_hss} for SVM with the \textit{linear} and \textit{rbf} kernels, respectively.
    %On the $x$-axis of these plots, the number of features utilized for evaluation is annotated, starting from the most important feature followed by incremental addition of the next best feature. The $y$-axis represents the mean TSS and HSS scores across all 24 methods.
    Looking at Fig.~\ref{fig:linear_mvts_vect_tss_hss}, it is evident that utilizing the \textit{linear} kernel results in consistent performance in terms of TSS and HSS, with any $k$ in the top-$k$ features. The insignificant fluctuations and the non-descending pattern in TSS and HSS demonstrate that the \textit{linear} kernel does not distinguish between different features' effectiveness, whether using the vectorized-based nor MTS-based FSS methods. Using the \textit{rbf} kernel (Fig.~\ref{fig:rbf_mvts_vect_tss_hss}), on the other hand, the performances are very distinct. The performance of the vectorized-based FSS methods drops drastically by adding the 7th feature. MTS-based methods, however, demonstrate higher tolerance for a larger number of top features. This observation might be of interest from the domain experts' point of view, as it allows a larger number of features to be picked for further investigations. An observation can also be made in favor of the vectorized-based methods; using those methods, with very few features (3 to 5) one can get similar or marginally better performance than using MTS-based methods with any (large or small) combination of the features. This is particularly important for operational forecast modules as they must take into account the factor of computation time. A model can perform faster on data with lower dimensionality.
    
    \begin{figure}
        \centering
        \centering\includegraphics[width=\linewidth]{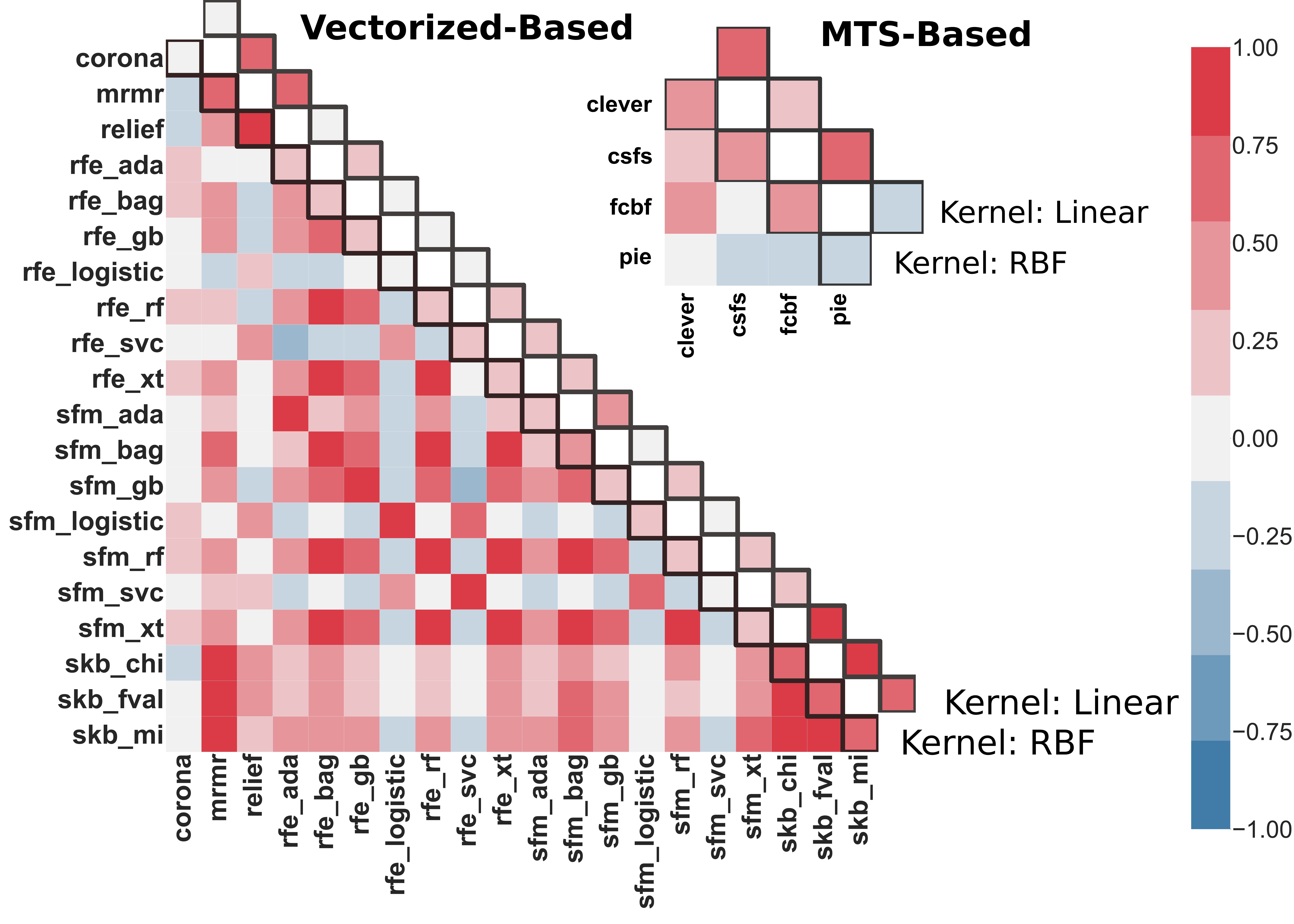}
        \caption{Cross comparison of the 24 FSS methods: each cell of the heatmaps (except those on the diagonal) represents the Pearson correlation of two rankings. The diagonal cells compares a ranking with that of SVM (univariate) using a \textit{linear} or \textit{rbf} kernel. The rankings of vectorized-based and MTS-based FSS methods are separated for a better visibility.}
        \label{fig:corr_heatmap}
        \vspace{-0.5cm}
    \end{figure}
    
    % Heatmap: Cross-comparison
    % Average Correlation | No Diagonal
    %   - RBF    --> Vect: 0.5980   MTS: 0.3910
    %   - LINEAR --> Vect: 0.5981   MTS: 0.3910
    % Average Correlation | Only Diagonal
    %   - RBF    --> Vect: 0.3066   MTS: 0.2952
    %   - LINEAR --> Vect: 0.2907   MTS: 0.3539

    The result of our cross comparison evaluation, is illustrated through the heatmaps in Fig.~\ref{fig:corr_heatmap}. 
    %ach cell of these heatmaps represents the Pearson correlation of two rankings returned by two FSS methods.
    Looking at the heatmaps, it is not difficult to see that the distribution of the correlations is negatively skewed (i.e., in favor of the positive values). Using the Kolmogorov-Smirnov test \cite{kolmogorovSmirnov1933sulla}, we assess whether the distribution of correlations is different from the uniform noise or not. With the test statistic of $\sim\!0.48$ associated with the very small p-value of $4.1\mathrm{e}{-122}$, we confidently reject the null hypothesis that the two distributions are similar. Therefore, the FSS methods, with the overall average correlation (using \textit{rbf}) of $\sim\!0.60$ for vectorized-based and $\sim\!0.39$ for MTS-based methods (non-diagonal cells), agree with each other much more than they disagree. On the degree of this agreement, the theoretical expectation might be to have a near-perfect average (linear) correlation.
    
    In a closer look, for vectorized-based FSS methods (the larger heatmap) a strong correlation ($\sim\!90\%$) is observed between methods with the same estimator irrespective of their different selection strategies. For example, \textit{rfe\_logistic} and \textit{sfm\_logistic} both use Logistic Regression as their estimator, but the former uses Recursive Feature Elimination (RFE) as its selection strategy while the latter uses Select From Model (SFM). While the computational complexity of RFE is significantly higher than that of SFM, on the SWAN-SF dataset they return similar feature rankings. This observation is not unique to those two methods only. A good correlation (60\%-80\%) between strategies having tree-based estimators is also observed, e.g., Random Forest (RF), Extra Trees (XT), Gradient Boosting (GB), Ada Boosting (ADA), and Bagging Trees (bag). We believe the main reason behind this similarity is their similar heuristics, and this should not be confused with the reliability of their results.
    
    For MTS-Based methods none of the utilized FSS methods demonstrate strong correlation within themselves. This can also be attributed to the inherent differences in the algorithmic approaches across all the MTS-based methods. That said, such FSS methods might be good candidates for making an ensemble feature selection model. This, however, is just a hypothesis yet to be investigated.
    
    Looking at the diagonals of the heatmaps, the average correlations with the SVM's rankings (using \textit{rbf}) are $0.31$ and $0.30$ for the vectorized-based and MTS-based methods, respectively. Interestingly, most of the FSS methods whose rankings are much more similar to those obtained by SVM are in fact filter methods, such as \textit{mRMR}, Relief, Select $k$-Best (with different estimators), and \textit{FBCF}. CLeVer which is an embedded FSS method also correlates very strongly with the SVM's ranking.
    
    \begin{figure}
        \centering
        \centering\includegraphics[width=\linewidth]{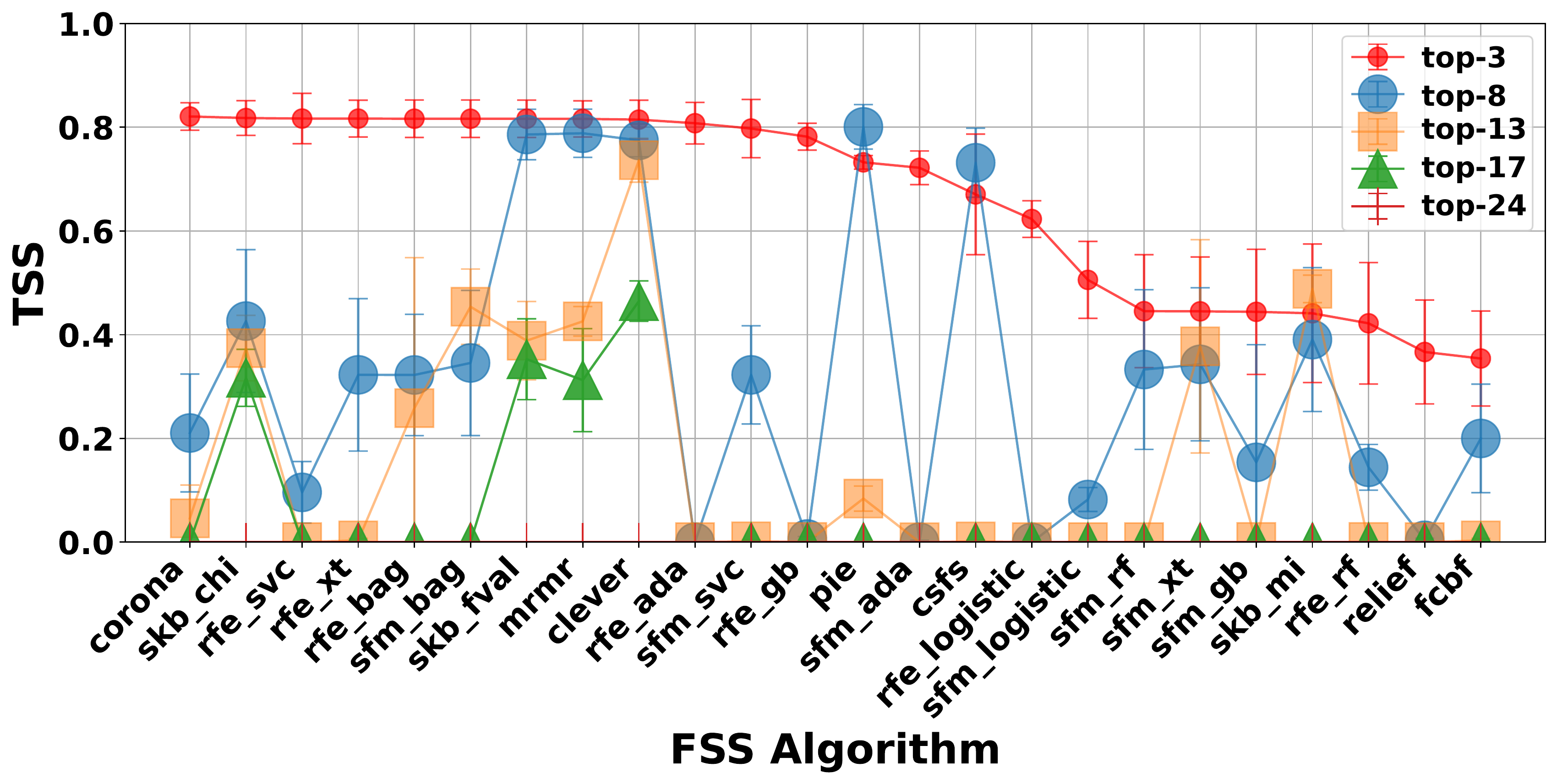}
        \caption{Performance of the top-k features per FSS method, reported in terms of TSS, using SVM with \textit{rbf} kernel. The $x$-axis represents the 24 FSS methods and their corresponding top-k features based on TSS score. The FSS methods on the $x$-axis are sorted by the TSS report of their top-3 features.}
        \label{fig:top_k_tss}
        \vspace{-0.4cm}
    \end{figure}
    
    \begin{figure}
        \centering
        \centering\includegraphics[width=\linewidth]{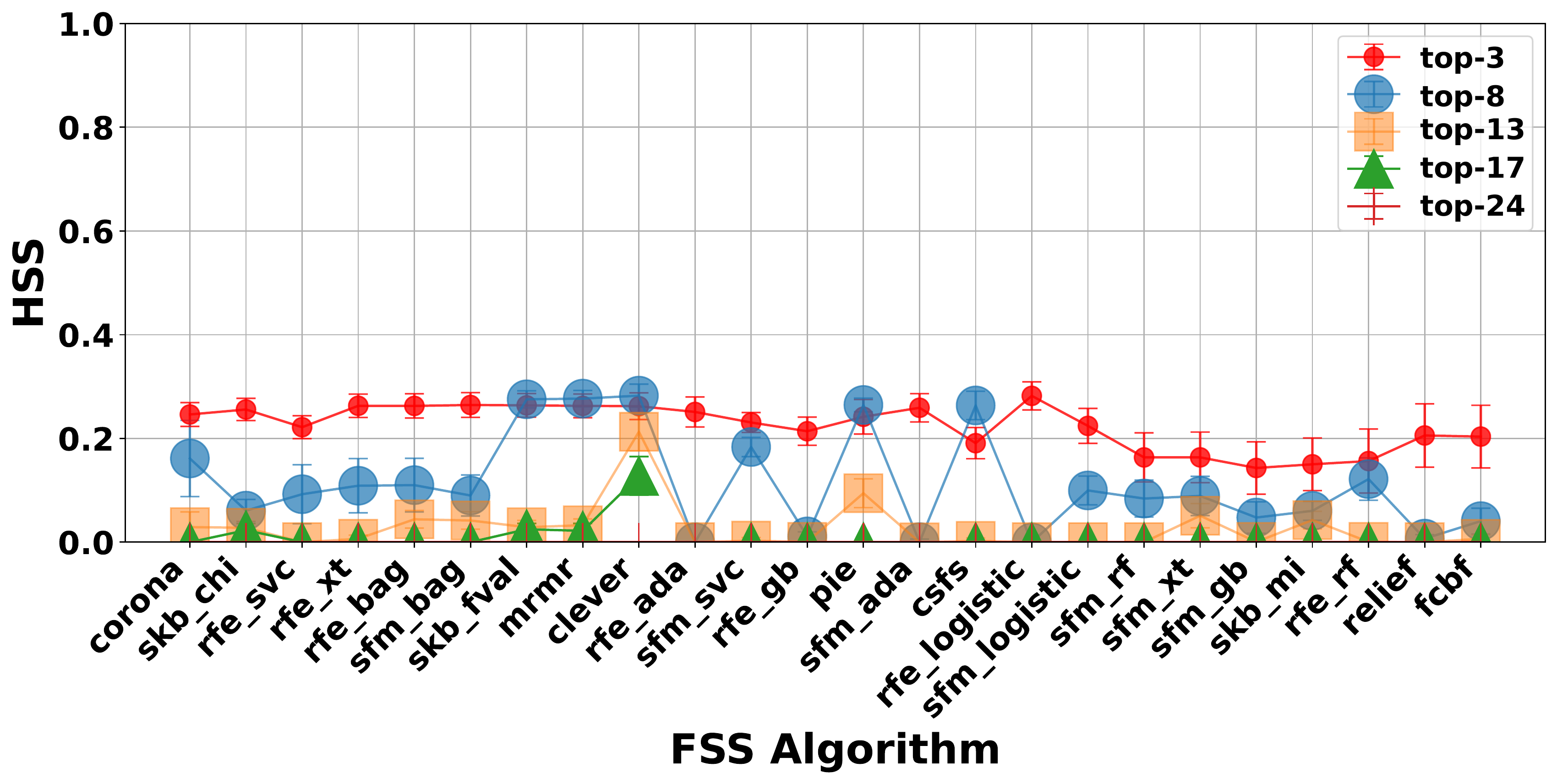}
        \caption{Performance of the top-k features per FSS method, reported in terms of HSS using SVM with \textit{rbf} kernel. The $x$-axis represents the 24 FSS methods and their corresponding top-k features based on HSS score. The FSS methods on the $x$-axis are sorted by the TSS (not HSS) report of their top-3 features.}
        \label{fig:top_k_hss}
        \vspace{-0.6cm}
    \end{figure}
    
    Another way to assess the collective relevance of the features is by looking at the performance SVM can achieve using the top-$k$ features. As illustrated in Figs.~\ref{fig:top_k_tss} and \ref{fig:top_k_hss}, most of the methods' top-3 features (first 11 ones) perform relatively well, in terms of TSS and HSS. But when more features are used together, i.e., top-17, most of the RFE-based methods (e.g., \textit{rfe\_ada}, \textit{rfe\_gb}, \textit{rfe\_logistic}) perform significantly worse. In contrast, (yet again) the filter methods seem to have selected their large set of features very well, as manifested by the TSS and HSS obtained by the top-17 features selected by \textit{skb\_chi}, \textit{skb\_fval}, and \textit{mRMR}. The top-17 features of the CLeVer method achieve the highest TSS and HSS compared to other methods in this group. CLeVer utilizes an unsupervised estimator ($k$-means) to drop the redundant features, which might be the reason for its selection to maintain such a high performance. The outperforming FSS methods with their top-8 features are also mainly filters, e.g., \textit{skb\_fval}, \textit{mRMR}, \textit{CSFS}, and \textit{PIE}.
    
    It is also interesting to note that although previously Corona did not show a high correlation with any of the other vectorized-based FSS methods (see Fig.~\ref{fig:corr_heatmap}), neither had it any correlation with the rankings obtained by SVM, its top-3 selection outperforms (in terms of TSS) all other top-$k$ features of all FSS methods. This does not hold true however, when HSS is used for evaluation.

% ----------------------------------
%
%       CONCLUSION
%
% ----------------------------------
\section{Conclusion}\label{sec:conclusion}
    In summary, our FSS investigation on SWAN-SF dataset brings us to the following conclusions:
    \begin{itemize}[leftmargin=*]
        \item Although different FSS methods rank SWAN-SF's features differently, we statistically showed that overall these methods agree with each other's rankings much more than they disagree. To avoid presenting an unreliable ranking of the features, we presented the FSS methods whose top-$k$ features performed better than others (see Figs.~\ref{fig:top_k_tss} and \ref{fig:top_k_hss}).
        % Using our publicly available code, and the freely available data, interested researchers should be able to reproduce our experiments and continue this work in their desired direction.
        \item The high correlations between FSS methods' returned rankings are often rooted in the similarities in their strategies and not necessarily the methods' confidence in their rankings. These similarities can be looked for in (1) the estimators (for wrapper and embedded methods), the statistical inference (for filter methods), and the embedded selection algorithms.
        \item In the absence of domain knowledge, the reliability of the ranked features cannot be trusted on its own, and it should be further tested by (1) cross-verification with the rankings obtained by other methods made up of non-similar components; (2) being verified by the same verification metrics that will be used for evaluation of the main problem. In this study, we used SVM as our control classifier, and TSS and HSS to keep track of its performance.
        \item Ambiguities in correlations between different FSS methods indicate that the performance of a solar flare forecast model (trained on SWAN-SF) may depend on the choice of the FSS method. Therefore, further investigation on a more appropriate FSS method is warranted.
        \item We observed that the top-$k$ features returned by FSS methods may perform differently, depending on the size of $k$. Those which work better with smaller values of $k$ are useful particularly for situations where the time and space complexity matters, whereas others can find a larger number of features without adding too much of redundancy. These are often more useful when the objective is to obtain domain insight into the features and their importance.
        \item All FSS methods (except Clever) utilized in this study are supervised methods, partially because there is a large emphasis on this group of methods in the literature. It was interesting, however, to observer that Clever outperformed most of the supervised methods. This hints at the importance of such methods and invites further investigation in that direction on SWAN-SF features.
    \end{itemize}

\section*{Acknowledgment}
    The authors would like to thank Heramb C. Lonkar and Egill Gunnarsson for kindly sharing their implementations of two FSS methods with us.
    
    This work was supported in part by two NASA Grant Awards [No. NNH14ZDA001N, 80NSSC20K1352], and two NSF Grant Awards [No. AC1443061 and AC1931555]. The AC1443061 award has been supported by funding from the Division of Advanced Cyber infrastructure within the Directorate for Computer and Information Science and Engineering, the Division of Astronomical Sciences within the Directorate for Mathematical and Physical Sciences, and the Division of Atmospheric and Geospace Sciences within the Directorate for Geosciences. VMS acknowledges the NSF FDSS grant 1936361 and NASA grant 80NSSC20K0302.

% \section*{References}

\bibliographystyle{./IEEEtran}
\bibliography{mybib}

\begin{thebibliography}{10}
\providecommand{\url}[1]{#1}
\csname url@samestyle\endcsname
\providecommand{\newblock}{\relax}
\providecommand{\bibinfo}[2]{#2}
\providecommand{\BIBentrySTDinterwordspacing}{\spaceskip=0pt\relax}
\providecommand{\BIBentryALTinterwordstretchfactor}{4}
\providecommand{\BIBentryALTinterwordspacing}{\spaceskip=\fontdimen2\font plus
\BIBentryALTinterwordstretchfactor\fontdimen3\font minus
  \fontdimen4\font\relax}
\providecommand{\BIBforeignlanguage}[2]{{%
\expandafter\ifx\csname l@#1\endcsname\relax
\typeout{** WARNING: IEEEtran.bst: No hyphenation pattern has been}%
\typeout{** loaded for the language `#1'. Using the pattern for}%
\typeout{** the default language instead.}%
\else
\language=\csname l@#1\endcsname
\fi
#2}}
\providecommand{\BIBdecl}{\relax}
\BIBdecl

\bibitem{eastwood2017economic}
\BIBentryALTinterwordspacing
J.~Eastwood, E.~Biffis, M.~Hapgood, L.~Green, M.~Bisi, R.~Bentley, R.~Wicks,
  L.-A. McKinnell, M.~Gibbs, and C.~Burnett, ``The economic impact of space
  weather: Where do we stand?'' \emph{Risk Analysis}, vol.~37, no.~2, pp.
  206--218, 2017. [Online]. Available: \url{https://doi.org/10.1111/risa.12765}
\BIBentrySTDinterwordspacing

\bibitem{nrc:spaceweather}
\BIBentryALTinterwordspacing
N.~R. Council, \emph{Severe Space Weather Events--Understanding Societal and
  Economic Impacts: A Workshop Report}.\hskip 1em plus 0.5em minus 0.4em\relax
  Washington, DC: The National Academies Press, 2008. [Online]. Available:
  \url{https://doi.org/10.17226/12507}
\BIBentrySTDinterwordspacing

\bibitem{barnes2016comparison}
G.~{Barnes}, K.~D. {Leka}, C.~J. {Schrijver}, T.~{Colak}, R.~{Qahwaji}, O.~W.
  {Ashamari}, Y.~{Yuan}, J.~{Zhang}, R.~T.~J. {McAteer}, D.~S. {Bloomfield},
  P.~A. {Higgins}, P.~T. {Gallagher}, D.~A. {Falconer}, M.~K. {Georgoulis},
  M.~S. {Wheatland}, C.~{Balch}, T.~{Dunn}, and E.~L. {Wagner}, ``A comparison
  of flare forecasting methods. i. results from the
  {\textquotedblleft}all-clear{\textquotedblright} workshop,'' \emph{apj}, vol.
  829, no.~2, p.~89, Oct. 2016.

\bibitem{bobra2015solar}
\BIBentryALTinterwordspacing
M.~G. Bobra and S.~Couvidat, ``Solar flare prediction using sdo/hmi vector
  magnetic field data with a machine-learning algorithm,'' \emph{The
  Astrophysical Journal}, vol. 798, no.~2, p. 135, 2015. [Online]. Available:
  \url{doi.org/10.1088/0004-637X/798/2/135}
\BIBentrySTDinterwordspacing

\bibitem{sadykov2017PIL}
V.~M. {Sadykov} and A.~G. {Kosovichev}, ``{Relationships between
  Characteristics of the Line-of-sight Magnetic Field and Solar Flare
  Forecasts},'' \emph{The Astrophysical Journal}, vol. 849, no.~2, p. 148, Nov.
  2017.

\bibitem{nishizuka2017flareforecast}
N.~{Nishizuka}, K.~{Sugiura}, Y.~{Kubo}, M.~{Den}, S.~{Watari}, and M.~{Ishii},
  ``{Solar Flare Prediction Model with Three Machine-learning Algorithms using
  Ultraviolet Brightening and Vector Magnetograms},'' \emph{The Astrophysical
  Journal}, vol. 835, no.~2, p. 156, Feb. 2017.

\bibitem{angryk2019multivariate}
\BIBentryALTinterwordspacing
R.~A. Angryk, P.~C. Martens, B.~Aydin, D.~Kempton, S.~S. Mahajan, S.~Basodi,
  A.~Ahmadzadeh, S.~F. Boubrahimi, S.~M. Hamdi, M.~A. Schuh \emph{et~al.},
  ``Multivariate time series dataset for space weather data analytics,''
  \emph{Sci. Data}, 2019. [Online]. Available:
  \url{https://doi.org/10.1038/s41597-020-0548-x}
\BIBentrySTDinterwordspacing

\bibitem{kohavi1997wrappers}
\BIBentryALTinterwordspacing
R.~Kohavi and G.~H. John, ``Wrappers for feature subset selection,''
  \emph{Artif. Intell.}, vol.~97, no. 1-2, pp. 273--324, 1997. [Online].
  Available: \url{https://doi.org/10.1016/S0004-3702(97)00043-X}
\BIBentrySTDinterwordspacing

\bibitem{guyon2003introduction}
I.~Guyon and A.~Elisseeff, ``An introduction to variable and feature
  selection,'' \emph{Journal of machine learning research}, vol.~3, no. Mar,
  pp. 1157--1182, 2003.

\bibitem{miao2016survey}
J.~Miao and L.~Niu, ``A survey on feature selection,'' \emph{Procedia Computer
  Science}, vol.~91, pp. 919--926, 2016.

\bibitem{lal2004support}
\BIBentryALTinterwordspacing
T.~N. Lal, M.~Schr{\"{o}}der, T.~Hinterberger, J.~Weston, M.~Bogdan,
  N.~Birbaumer, and B.~Sch{\"{o}}lkopf, ``Support vector channel selection in
  {BCI},'' \emph{{IEEE} Trans. Biomed. Eng.}, vol.~51, no.~6, pp. 1003--1010,
  2004. [Online]. Available: \url{https://doi.org/10.1109/TBME.2004.827827}
\BIBentrySTDinterwordspacing

\bibitem{guyon2002gene}
I.~Guyon, J.~Weston, S.~Barnhill, and V.~Vapnik, ``Gene selection for cancer
  classification using support vector machines,'' \emph{Machine learning},
  vol.~46, no.~1, pp. 389--422, 2002.

\bibitem{bishop1995neural}
C.~Bishop, ``Neural networks for pattern recognition,'' 1995.

\bibitem{yang2005supervised}
K.~Yang, H.~Yoon, and C.~Shahabi, ``A supervised feature subset selection
  technique for multivariate time series,'' in \emph{Proceedings of the
  workshop on feature selection for data mining: Interfacing machine learning
  with statistics}, 2005, pp. 92--101.

\bibitem{smialowski2010pitfalls}
\BIBentryALTinterwordspacing
P.~Smialowski, D.~Frishman, and S.~Kramer, ``Pitfalls of supervised feature
  selection,'' \emph{Bioinform.}, vol.~26, no.~3, pp. 440--443, 2010. [Online].
  Available: \url{https://doi.org/10.1093/bioinformatics/btp621}
\BIBentrySTDinterwordspacing

\bibitem{Yoon2005FeatureSS}
H.~Yoon, K.~Yang, and C.~Shahabi, ``Feature subset selection and feature
  ranking for multivariate time series,'' \emph{IEEE Transactions on Knowledge
  and Data Engineering}, vol.~17, pp. 1186--1198, 2005.

\bibitem{robnik2003theoretical}
M.~Robnik-{\v{S}}ikonja and I.~Kononenko, ``Theoretical and empirical analysis
  of relieff and rrelieff,'' \emph{Machine learning}, vol.~53, no.~1, pp.
  23--69, 2003.

\bibitem{urbanowicz2018relief}
\BIBentryALTinterwordspacing
R.~J. Urbanowicz, M.~Meeker, W.~G.~L. Cava, R.~S. Olson, and J.~H. Moore,
  ``Relief-based feature selection: Introduction and review,'' \emph{J. Biomed.
  Informatics}, vol.~85, pp. 189--203, 2018. [Online]. Available:
  \url{https://doi.org/10.1016/j.jbi.2018.07.014}
\BIBentrySTDinterwordspacing

\bibitem{peng2005mrmr}
\BIBentryALTinterwordspacing
H.~Peng, F.~Long, and C.~H.~Q. Ding, ``Feature selection based on mutual
  information: Criteria of max-dependency, max-relevance, and min-redundancy,''
  \emph{{IEEE} Trans. Pattern Anal. Mach. Intell.}, vol.~27, no.~8, pp.
  1226--1238, 2005. [Online]. Available:
  \url{https://doi.org/10.1109/TPAMI.2005.159}
\BIBentrySTDinterwordspacing

\bibitem{ramirez2017fast}
\BIBentryALTinterwordspacing
S.~Ram{\'{\i}}rez{-}Gallego, I.~Lastra, D.~Mart{\'{\i}}nez{-}Rego,
  V.~Bol{\'{o}}n{-}Canedo, J.~M. Ben{\'{\i}}tez, F.~Herrera, and
  A.~Alonso{-}Betanzos, ``Fast-mrmr: Fast minimum redundancy maximum relevance
  algorithm for high-dimensional big data,'' \emph{Int. J. Intell. Syst.},
  vol.~32, no.~2, pp. 134--152, 2017. [Online]. Available:
  \url{https://doi.org/10.1002/int.21833}
\BIBentrySTDinterwordspacing

\bibitem{akadi2011two}
\BIBentryALTinterwordspacing
A.~E. Akadi, A.~Amine, A.~E. Ouardighi, and D.~Aboutajdine, ``A two-stage gene
  selection scheme utilizing {MRMR} filter and {GA} wrapper,'' \emph{Knowl.
  Inf. Syst.}, vol.~26, no.~3, pp. 487--500, 2011. [Online]. Available:
  \url{https://doi.org/10.1007/s10115-010-0288-x}
\BIBentrySTDinterwordspacing

\bibitem{zhang2007two}
\BIBentryALTinterwordspacing
Y.~Zhang, C.~H.~Q. Ding, and T.~Li, ``A two-stage gene selection algorithm by
  combining relieff and mrmr,'' in \emph{Proceedings of the 7th {IEEE}
  International Conference on Bioinformatics and Bioengineering, {BIBE} 2007,
  October 14-17, 2007, Harvard Medical School, Boston, MA, {USA}}.\hskip 1em
  plus 0.5em minus 0.4em\relax {IEEE} Computer Society, 2007, pp. 164--171.
  [Online]. Available: \url{https://doi.org/10.1109/BIBE.2007.4375560}
\BIBentrySTDinterwordspacing

\bibitem{ding2009feature}
S.~Ding, ``Feature selection based f-score and aco algorithm in support vector
  machine,'' in \emph{2009 Second International Symposium on Knowledge
  Acquisition and Modeling}, vol.~1, 2009, pp. 19--23.

\bibitem{vinh2012novel}
\BIBentryALTinterwordspacing
L.~T. Vinh, S.~Lee, Y.~Park, and B.~J. d'Auriol, ``A novel feature selection
  method based on normalized mutual information,'' \emph{Appl. Intell.},
  vol.~37, no.~1, pp. 100--120, 2012. [Online]. Available:
  \url{https://doi.org/10.1007/s10489-011-0315-y}
\BIBentrySTDinterwordspacing

\bibitem{azhagusundari2013feature}
B.~Azhagusundari, A.~S. Thanamani \emph{et~al.}, ``Feature selection based on
  information gain,'' \emph{International Journal of Innovative Technology and
  Exploring Engineering (IJITEE)}, vol.~2, no.~2, pp. 18--21, 2013.

\bibitem{yu2003feature}
\BIBentryALTinterwordspacing
L.~Yu and H.~Liu, ``Feature selection for high-dimensional data: {A} fast
  correlation-based filter solution,'' in \emph{Machine Learning, Proceedings
  of the Twentieth International Conference {(ICML} 2003), August 21-24, 2003,
  Washington, DC, {USA}}, T.~Fawcett and N.~Mishra, Eds.\hskip 1em plus 0.5em
  minus 0.4em\relax {AAAI} Press, 2003, pp. 856--863. [Online]. Available:
  \url{http://www.aaai.org/Library/ICML/2003/icml03-111.php}
\BIBentrySTDinterwordspacing

\bibitem{cinto2020extremegrad}
T.~{Cinto}, A.~L.~S. {Gradvohl}, G.~P. {Coelho}, and A.~E.~A. {da Silva},
  ``{Solar Flare Forecasting Using Time Series and Extreme Gradient Boosting
  Ensembles},'' \emph{Solar Physics}, vol. 295, no.~7, p.~93, Jul. 2020.

\bibitem{breiman2001randomforest}
L.~{Breiman}, ``{Random Forests.}'' \emph{Machine Learning}, vol.~45, pp.
  5--32, Jan. 2001.

\bibitem{liu2017rfflare}
C.~{Liu}, N.~{Deng}, J.~T.~L. {Wang}, and H.~{Wang}, ``{Predicting Solar Flares
  Using SDO/HMI Vector Magnetic Data Products and the Random Forest
  Algorithm},'' \emph{The Astrophysical Journal}, vol. 843, no.~2, p. 104, Jul.
  2017.

\bibitem{wang2020kernelPCA}
J.~{Wang}, Y.~{Zhang}, S.~A. {Hess Webber}, S.~{Liu}, X.~{Meng}, and T.~{Wang},
  ``{Solar Flare Predictive Features Derived from Polarity Inversion Line Masks
  in Active Regions Using an Unsupervised Machine Learning Algorithm},''
  \emph{The Astrophysical Journal}, vol. 892, no.~2, p. 140, Apr. 2020.

\bibitem{ahmed2013asap}
O.~W. {Ahmed}, R.~{Qahwaji}, T.~{Colak}, P.~A. {Higgins}, P.~T. {Gallagher},
  and D.~S. {Bloomfield}, ``{Solar Flare Prediction Using Advanced Feature
  Extraction, Machine Learning, and Feature Selection},'' \emph{Solar Physics},
  vol. 283, no.~1, pp. 157--175, Mar. 2013.

\bibitem{pearson1901liii}
K.~Pearson, ``Liii. on lines and planes of closest fit to systems of points in
  space,'' \emph{The London, Edinburgh, and Dublin philosophical magazine and
  journal of science}, vol.~2, no.~11, pp. 559--572, 1901.

\bibitem{wold1987principal}
S.~Wold, K.~Esbensen, and P.~Geladi, ``Principal component analysis,''
  \emph{Chemometrics and intelligent laboratory systems}, vol.~2, no. 1-3, pp.
  37--52, 1987.

\bibitem{zhao2019maximum}
Z.~Zhao, R.~Anand, and M.~Wang, ``Maximum relevance and minimum redundancy
  feature selection methods for a marketing machine learning platform,'' in
  \emph{2019 IEEE International Conference on Data Science and Advanced
  Analytics (DSAA)}.\hskip 1em plus 0.5em minus 0.4em\relax IEEE, 2019, pp.
  442--452.

\bibitem{scikit-learn}
F.~Pedregosa, G.~Varoquaux, A.~Gramfort, V.~Michel, B.~Thirion, O.~Grisel,
  M.~Blondel, P.~Prettenhofer, R.~Weiss, V.~Dubourg, J.~Vanderplas, A.~Passos,
  D.~Cournapeau, M.~Brucher, M.~Perrot, and E.~Duchesnay, ``Scikit-learn:
  Machine learning in {P}ython,'' \emph{Journal of Machine Learning Research},
  vol.~12, pp. 2825--2830, 2011.

\bibitem{han2005supervised}
\BIBentryALTinterwordspacing
S.~Han and A.~Niculescu{-}Mizil, ``Supervised feature subset selection and
  feature ranking for multivariate time series without feature extraction,''
  \emph{CoRR}, vol. abs/2005.00259, 2020. [Online]. Available:
  \url{https://arxiv.org/abs/2005.00259}
\BIBentrySTDinterwordspacing

\bibitem{Serra2014journal}
J.~Serra and J.~L. Arcos, ``An empirical evaluation of similarity measures for
  time series classification,'' \emph{Knowledge-Based Systems}, vol.~67, pp.
  305--314, 2014.

\bibitem{shi2000normalized}
J.~Shi and J.~Malik, ``Normalized cuts and image segmentation,'' \emph{IEEE
  Transactions on pattern analysis and machine intelligence}, vol.~22, no.~8,
  pp. 888--905, 2000.

\bibitem{HAN201329}
\BIBentryALTinterwordspacing
M.~Han and X.~Liu, ``Feature selection techniques with class separability for
  multivariate time series,'' \emph{Neurocomputing}, vol. 110, pp. 29--34,
  2013. [Online]. Available:
  \url{https://www.sciencedirect.com/science/article/pii/S0925231212009009}
\BIBentrySTDinterwordspacing

\bibitem{DVN/EBCFKM_2020}
\BIBentryALTinterwordspacing
R.~Angryk, P.~Martens, B.~Aydin, D.~Kempton, S.~Mahajan, S.~Basodi,
  A.~Ahmadzadeh, X.~Cai, S.~Filali~Boubrahimi, S.~M. Hamdi, M.~Schuh, and
  M.~Georgoulis, ``{SWAN-SF},'' 2020. [Online]. Available:
  \url{https://doi.org/10.7910/DVN/EBCFKM}
\BIBentrySTDinterwordspacing

\bibitem{hanssen1965relationship}
A.~Hanssen and W.~Kuipers, \emph{On the Relationship Between the Frequency of
  Rain and Various Meteorological Parameters (with Reference to the Problem of
  Objective Forecasting)}.\hskip 1em plus 0.5em minus 0.4em\relax Koninklijk
  Nederlands Meteorologisch Instituut, 1965, vol.~81.

\bibitem{balch2008updated}
\BIBentryALTinterwordspacing
C.~C. {Balch}, ``{Updated verification of the Space Weather Prediction Center's
  solar energetic particle prediction model},'' \emph{Space Weather}, vol.~6,
  no.~1, p. S01001, Jan. 2008. [Online]. Available:
  \url{doi.org/10.1029/2007SW000337}
\BIBentrySTDinterwordspacing

\bibitem{barnes2008evaluating}
\BIBentryALTinterwordspacing
G.~Barnes and K.~Leka, ``Evaluating the performance of solar flare forecasting
  methods,'' \emph{apjl}, vol. 688, no.~2, p. L107, 2008. [Online]. Available:
  \url{https://doi.org/10.1086/595550}
\BIBentrySTDinterwordspacing

\bibitem{Bloomfield_2012}
\BIBentryALTinterwordspacing
D.~S. {Bloomfield}, P.~A. {Higgins}, R.~T.~J. {McAteer}, and P.~T. {Gallagher},
  ``{Toward Reliable Benchmarking of Solar Flare Forecasting Methods},''
  \emph{apjl}, vol. 747, no.~2, p. L41, Mar. 2012. [Online]. Available:
  \url{https://iopscience.iop.org/article/10.1088/2041-8205/747/2/L41}
\BIBentrySTDinterwordspacing

\bibitem{ahmadzadeh2021how}
\BIBentryALTinterwordspacing
A.~Ahmadzadeh, B.~Aydin, M.~K. Georgoulis, D.~J. Kempton, S.~S. Mahajan, and
  R.~A. Angryk, ``How to train your flare prediction model: Revisiting robust
  sampling of rare events,'' \emph{The Astrophysical Journal Supplement
  Series}, vol. 254, no.~2, p.~23, May 2021. [Online]. Available:
  \url{http://dx.doi.org/10.3847/1538-4365/abec88}
\BIBentrySTDinterwordspacing

\bibitem{hostetter2019understanding}
M.~Hostetter, A.~Ahmadzadeh, B.~Aydin, M.~K. Georgoulis, D.~J. Kempton, and
  R.~A. Angryk, ``Understanding the impact of statistical time series features
  for flare prediction analysis,'' in \emph{2019 IEEE International Conference
  on Big Data (Big Data)}, 2019, pp. 4960--4966.

\bibitem{ahmadzadeh2019rare}
A.~Ahmadzadeh, B.~Aydin, D.~J. Kempton, M.~Hostetter, R.~A. Angryk, M.~K.
  Georgoulis, and S.~S. Mahajan, ``Rare-event time series prediction: A case
  study of solar flare forecasting,'' in \emph{2019 18th IEEE International
  Conference On Machine Learning And Applications (ICMLA)}, 2019, pp.
  1814--1820.

\bibitem{mvts2020ahmadzadeh}
\BIBentryALTinterwordspacing
A.~Ahmadzadeh, K.~Sinha, B.~Aydin, and R.~A. Angryk, ``Mvts-data toolkit: {A}
  python package for preprocessing multivariate time series data,''
  \emph{SoftwareX}, vol.~12, p. 100518, 2020. [Online]. Available:
  \url{https://doi.org/10.1016/j.softx.2020.100518}
\BIBentrySTDinterwordspacing

\bibitem{tavenard2020tslearn}
\BIBentryALTinterwordspacing
R.~Tavenard, J.~Faouzi, G.~Vandewiele, F.~Divo, G.~Androz, C.~Holtz, M.~Payne,
  R.~Yurchak, M.~Ru{\ss}wurm, K.~Kolar, and E.~Woods, ``Tslearn, {A} machine
  learning toolkit for time series data,'' \emph{J. Mach. Learn. Res.},
  vol.~21, pp. 118:1--118:6, 2020. [Online]. Available:
  \url{http://jmlr.org/papers/v21/20-091.html}
\BIBentrySTDinterwordspacing

\bibitem{kubat1997addressing}
M.~Kubat and S.~Matwin, ``Addressing the curse of imbalanced training sets:
  One-sided selection,'' in \emph{Proceedings of the Fourteenth International
  Conference on Machine Learning {(ICML} 1997), Nashville, Tennessee, USA, July
  8-12, 1997}, D.~H. Fisher, Ed.\hskip 1em plus 0.5em minus 0.4em\relax Morgan
  Kaufmann, 1997, pp. 179--186.

\bibitem{kolmogorovSmirnov1933sulla}
A.~Kolmogorov-Smirnov, A.~Kolmogorov, and M.~Kolmogorov, ``Sulla determinazione
  emp{\'i}rica di uma legge di distribuzione,'' 1933.

\end{thebibliography}

\end{document}